\documentclass{aa}  
\usepackage[colorlinks=true,linkcolor=blue,citecolor=blue]{hyperref}%
\usepackage{graphicx}
\usepackage{txfonts}
\usepackage{amsmath}
\usepackage{adjustbox}
\usepackage{placeins}
\usepackage{float}
\usepackage{etoolbox}

\newif\ifusebold

\useboldfalse

\ifusebold
   \newcommand{\mybold}[1]{\textbf{#1}}
\else
   \newcommand{\mybold}[1]{#1}
\fi

\begin{document}

   \title{The Impact of Star Formation Histories on the Inner Dark Matter Density Slopes of Galaxies}

   \author{J. Sarrato-Alós
          \inst{1}\fnmsep\inst{2},
          J. Bullock
          \inst{3},
          A. Di Cintio \inst{2}\fnmsep\inst{1},
          C. Brook \inst{2}\fnmsep\inst{1},
          F. Valenciano\inst{1}\fnmsep\inst{2} \&
          A. V. Macciò\inst{4}\fnmsep\inst{5}
         }

   \institute{Instituto de Astrofísica de Canarias (IAC), Calle Via Láctea s/n, E-38205 La Laguna, Tenerife, Spain\\
   \email{jorge.sarrato@iac.es}
         \and
             Universidad de La Laguna, Avda. Astrofísico Fco. Sánchez s/n, E-38206 La Laguna, Tenerife, Spain
         \and
              Department of Physics \& Astronomy, University of Southern California, Los Angeles, CA 90089, USA
         \and
             New York University Abu Dhabi, PO Box 129188, Abu Dhabi, United Arab Emirates
         \and
             Center for Astrophysics and Space Science, New York University Abu Dhabi, Abu Dhabi, PO Box 129188, Abu Dhabi, UAE
             }

   \date{Received x xx, xxxx; accepted x xx, xxxx}

 
  \abstract
   {}
   {We aim to investigate the connection between star formation histories (SFHs) and the inner dark matter density profiles of simulated galaxies. In particular, we test whether the burstiness and temporal distribution of star formation influence the formation of cored versus cuspy dark matter profiles.}
   {We homogeneously analysed simulated galaxies from the NIHAO and FIRE-2 projects. For each galaxy, we derived dark matter density profiles and measured the logarithmic slope in the inner region of the dark matter halo (1–2\% of R$_{\rm vir}$). To characterise star formation burstiness, we introduced a criterion based on comparing the star formation rate (SFR) averaged over two distinct timescales. We further quantified the \mybold{temporal concentration} of SFHs by computing $M_{\star, \rm post}$ / $M_{\star, \rm pre}$, the ratio of stellar mass formed after versus before the epoch of reionisation at redshift z $\sim$ 6.5.}
   {Homogeneous analysis reveals that inner slope versus stellar-to-halo mass ratio trends for NIHAO and FIRE-2 galaxies are in much better agreement than reported in previous works. The burstiness and \mybold{post-to-pre reionisation stellar mass ratio} explain the scatter in the inner slope versus stellar-to-halo mass ratio relation, revealing that galaxies with above average burstiness and more extended SFHs are more efficient at developing cored dark matter profiles. In contrast, galaxies with smoother SFHs and earlier stellar mass assembly tend to maintain cuspier dark matter profiles. We present an analytic expression that improves predictions for the inner slope using the parameter $M_{\star \rm,post}$ / $M_{\star \rm,pre}$, which reduces the mean squared error in both simulation suites relative to previous formulations based solely on the stellar-to-halo mass ratio.}
   {}


   \titlerunning{The Impact of SFHs on the Inner Slope}
   \authorrunning{J. Sarrato-Alós et al.}

   \maketitle

%

\section{Introduction}

Understanding how baryonic processes influence the distribution of dark matter in galaxies is a key objective in the study of galaxy formation and evolution. One of the most studied topics in this area is the discrepancy between the central dark matter profiles predicted by simulations and those inferred from observations of dwarf and low surface brightness galaxies. While cold dark matter only simulations robustly predict cuspy inner profiles \citep{NFW}, observational studies often find shallower, core-like distributions \citep{Blok_obs_cores, Oh_2011}. This "core-cusp" tension has motivated investigation into the potential for baryonic physics (particularly stellar feedback) to modify the dark matter structure found in cold dark matter only simulations, especially in the central regions of galaxies.

Numerous theoretical and numerical works have demonstrated that star formation and feedback processes can significantly impact the inner slope of dark matter halos. In particular, the total amount of energy injected by stellar feedback, relative to the total halo mass of the galaxy, plays a critical role in driving dark matter core formation. This property is closely correlated with the stellar-to-halo mass ratio \citep{penarrubia_energy_injection,Pontzen,brook_sn}. Simulations show that repeated gas outflows associated with star formation can generate potential fluctuations that dynamically heat the dark matter, pushing it outward and flattening the central cusp \citep{dicintio_profile, dicintio_mstar_mhalo,Tollet, Lazar}. In particular, \cite{dicintio_mstar_mhalo} fitted the relationship between the dark matter profile inner slope and the stellar-to-halo mass ratio using Eq.~\ref{eq:dicintio}, whilst \cite{Tollet} and \cite{Lazar} used the more complex Eq.~\ref{eq:tollet}, with $x=M_{\star}/M_{\rm halo}$ in both cases. It is worth noting that different studies adopt different halo mass definitions: some compute the halo mass $M_{\rm halo}$ using an overdensity of 200 times the critical density of the universe ($M_{\rm 200c}$), whereas others use the virial overdensity ($M_{\rm vir}$). Equally, $R_{\rm halo}$ is defined as the radius that encloses a total mass of $M_{\rm halo}$ according to the adopted overdensity. These studies have revealed that core formation is most efficient at intermediate stellar-to-halo mass ratios ($M_{\star}/M_{\rm halo}$ $\sim$ 0.005-0.007), providing a useful first order predictor of halo structure.

\begin{equation}
\label{eq:dicintio}
    \left.\frac{d\log{\rho}}{d\log{r}}\right|_{r = 0.015 R_{\rm halo}} (x) = n - \log_{10} \left[ \left( \frac{x}{x_{0}}\right)^{-\beta} + \left( \frac{x}{x_{0}}\right)^\gamma \right]
\end{equation}

\begin{equation}
\label{eq:tollet}
    \left.\frac{d\log{\rho}}{d\log{r}}\right|_{r = 0.015 R_{\rm halo}} (x) = n - \log_{10} \left[ n_{1} \left( 1+\frac{x}{x_{1}}\right)^{-\beta} + \left( \frac{x}{x_{0}}\right)^\gamma \right]
\end{equation}

However, while these results emphasise the importance of integrated stellar feedback over cosmic time, the detailed temporal structure of star formation (the extent to which it is bursty or continuous, or how \mybold{much stellar mass is formed before versus after a given time}) may further modulate the effectiveness of this process. Specifically, it has been proposed that bursty star formation, which drives stronger and more abrupt changes in the gravitational potential, may be more effective at reshaping dark matter distributions than smoother, extended star formation histories with similar total stellar mass \citep{Navarro_1996, read_mass_loss, Pontzen_2014}. \cite{dicintio_2017} demonstrated that both the burstiness and overall duration of star formation are key drivers of the structural differences between diffuse and compact galaxies. Systems that experience extended, highly bursty SFHs tend to develop more extended dark matter cores, whereas those with shorter, less bursty SFHs exhibit cuspier central dark matter profiles and smaller effective radii (see their Fig. 4). This motivates the need to quantify burstiness as a distinct parameter, beyond stellar-to-halo mass ratio, in understanding dark matter core formation.

Apart from burstiness, \cite{Muni_2024} highlighted the importance of the \mybold{temporal concentration} of the SFH on the process of core formation. Specifically, they found the ratio of stellar mass that formed after and before the reionisation epoch \mybold{(hereafter referred to as post-to-pre reionisation stellar mass ratio)} shows better correlation with the inner dark matter density of dwarf galaxies than the stellar-to-halo mass ratio in the EDGE simulations \citep{edge1, edge2}.

Observational and theoretical efforts have begun to characterise bursty star formation in low mass galaxies \citep{Weisz_burst, Emami_burst}, and cosmological simulations increasingly resolve these time variable feedback processes. However, the relationship between star formation burstiness and dark matter structure has yet to be systematically quantified. In particular, it remains unclear whether galaxies with similar stellar-to-halo mass ratios but different star formation histories exhibit systematically different inner density profiles, and whether burstiness can account for the scatter in the observed relation between the $M_{\star}/M_{\rm halo}$ ratio and the inner slope.

In this paper, we address this gap by homogeneously analysing two suites of simulated galaxies to quantify the burstiness of their SFHs and measure the corresponding slopes of their dark matter density profiles. We show that galaxies with similar stellar mass but differing degrees of burstiness can present remarkably different inner dark matter structures, with more bursty galaxies tending to produce cored profiles more often. Our results suggest that burstiness introduces an important second order effect that can help explain the diversity in inner halo slopes. Additionally, we find the \mybold{post-to-pre reionisation stellar mass ratio} has a similar effect to that of burstiness on the inner slope of the dark matter profile. By explicitly linking time resolved star formation behaviour to dark matter core formation, this work offers new insight into the role of baryonic feedback in shaping galaxy structure.

The paper is organised as follows. Section \ref{sec:simulations} introduces the two suites of simulated galaxies that form the basis of our analysis. In Section \ref{sec:results}, we present the main results from the homogeneous analysis of both suites. This section is structured to first address the relationship between the burstiness metric and the inner dark matter profile slopes (Section \ref{sec:results-burst}). We then investigate the correlation between the \mybold{post-to-pre reionisation stellar mass ratio} and the inner dark matter slopes in Section \ref{sec:results-mpostpre}. Section \ref{sec:secondorderfit} reports an improved fitting equation for the inner slope that combines the effects of the stellar-to-halo mass ratio and the  \mybold{post-to-pre reionisation stellar mass ratio}. Finally, our conclusions are summarised in Section \ref{sec:conclusions}.

\section{Simulations}
\label{sec:simulations}

\subsection{The NIHAO project}
\label{sec:NIHAO}

The NIHAO project (Numerical Investigations of Hundred Astrophysical Objects) is a suite of high resolution cosmological zoom-in hydrodynamical simulations based on the {\sc gasoline2} code \citep{wadsley2017} and first presented in \cite{wang2015}. The NIHAO project adopts Planck cosmology \citep{planckcosmo}, using the following parameters: H$_{\text{0}}$ = 100h km s$^{\text{-1}}$ Mpc$^{\text{-1}}$ with h = 0.671, $\Omega_{\text{m}}$ = 0.3175, $\Omega_{\Lambda}$ = 0.6824, $\Omega_{\text{b}}$ = 0.049, and $\sigma_{\text{8}}$ = 0.8344 \citep{wang2015}.

NIHAO simulations include all fundamental processes involved in galaxy formation. More precisely the implementation of star formation and stellar feedback mechanisms follows the model established in the Making Galaxies In a Cosmological Context \citep[MaGICC;][]{stinson13} simulations. This particular framework has proved successful in reproducing a broad range of observed galaxy scaling relations \citep{brook12b,maccio2012}. \mybold{The suite correctly reproduces the observed stellar mass-halo mass relation and generates a diverse range of realistic morphologies, from irregular dwarfs to well-defined stellar discs in Milky Way-mass systems \citep{wang2015}. Furthermore, its galaxies align well with the SPARC sample \citep{Santossantos2018} and the Tully-Fisher relation \citep{Dutton2017}.} Star formation is allowed \mybold{according to the Kennicutt–Schmidt Law} in dense gas that \mybold{satisfies a temperature of T < $1.5\times10^{4}$ K and} exceeds a density threshold number, set at $n_{\rm th} = 10.3$ cm$^{-3}$, using an initial mass function (IMF) from \cite{chabrier03}.

\mybold{Stellar feedback is implemented in two distinct epochs. The first involves early stellar feedback from massive stars prior to their supernova explosions \citep{stinson13}. In the NIHAO suite, this consists of injecting 13\% of the total stellar flux (equivalent to $2 \times 10^{50}$ erg of thermal energy per M$_{\odot}$ of the entire stellar population) into the surrounding gas while leaving radiative cooling enabled.} \mybold{The second epoch consists of supernova feedback from stars in the mass range of $8~- 40~M_{\odot}$, which are modelled as collective effects at larger sub-grid scales (typically $\sim 100$ pc to 1 kpc).} This feedback is modelled using a \mybold{thermal} blast-wave approach \citep{stinson06}. \mybold{Because the gas receiving the energy is dense, cooling is artificially delayed for particles inside the blast region for $\sim 30$ Myr to prevent numerical over-cooling.}

Gas evolution includes metal-line cooling, photoionisation, and ultraviolet heating, based on the prescriptions detailed by \cite{shen}. Specifically, the ultraviolet background is implemented via the \cite{haardtmadau} model, which leads to the complete ionisation of hydrogen in the intergalactic medium by z $\sim$ 6.7.

The mass and spatial resolution of the simulations allow the inner structure of galaxies to be resolved down to below 1\% of the virial radius \citep[e.g.][]{Tollet}. Half-light radii are well captured, with spatial resolutions ranging from about 100 pc in low-mass systems to 800 pc in the most massive galaxies. \mybold{Individual particle masses scale with the total mass of the simulated galaxy, spanning $6\times10^{2}$ to $3\times10^{5}$ M$_{\odot}$ for baryons and $3\times10^{3}$ to $2\times10^{6}$ M$_{\odot}$ for dark matter in the highest-resolution zoom-in region. The simulation achieves high temporal precision, with star formation computed at intervals of 0.84 Myr \citep{dutton2019}; meanwhile, the maximum time-step for force calculations is 13.5 Myr, with adaptive refinement allowing for steps as small as 12.9 yr depending on local particle acceleration.}

We used Amiga Halo Finder \citep[AHF][]{ahf} and selected the main isolated halo from each zoom-in simulation as long as it contained at least one hundred star particles, resulting in a sample of 93 galaxies.

\subsection{The FIRE project}
\label{sec:FIRE}

The FIRE project comprises several sets of cosmological zoom-in simulations generated with the GIZMO \citep{gizmo} code, using slightly different cosmological parameters depending on the run. Some follow the Planck cosmology \citep{planckcosmo}, while others adopt parameters from the Assembling Galaxies Of Resolved Anatomy \citep[AGORA;][]{AGORA_cosmo} project: H$_{\text{0}}$ = 100h km s$^{\text{-1}}$ Mpc$^{\text{-1}}$ with h = 0.702, $\Omega_{\text{m}}$ = 0.272, $\Omega_{\Lambda}$ = 0.728, $\Omega_{\text{b}}$ = 0.0455, and $\sigma_{\text{8}}$ = 0.807.

Specifically, we use the FIRE-2 simulations \citep{Hopkins_2018}, which include a detailed model of galaxy formation physics. \mybold{These simulations produce a realistic diversity of Milky Way-mass morphologies, ranging from thin, rotationally supported stellar discs to compact, bulge-dominated systems \citep{GarrisonKimmel_2018}. Furthermore, the model matches the observed structural scaling relations of Local Group dwarf galaxies, including their stellar half-mass radii and velocity dispersions \citep{Fitts_2017}. FIRE-2 is noted for capturing the multiphase interstellar medium and generating realistic star formation histories and metallicity gradients without parameter fine-tuning. It naturally produces the low surface brightness and large effective radii characteristic of local dwarf populations through stellar feedback-driven galactic expansion \citep{Chan_2018}. While the simulations match observed scaling relations for angular momentum and the Tully-Fisher relation, gas in low-mass systems is found to be more dispersion-supported than typical observed galaxies \citep{ElBadry_2018a, ElBadry_2018b}. On top of a density threshold of $n_{\rm th} = 1000$ cm$^{-3}$, star formation is restricted to occur only when gas is self-gravitating, self-shielding and Jeans unstable. Once a star particle forms, it is treated as a single stellar population with an age, metallicity, and mass inherited from its progenitor gas particle. Feedback quantities are tabulated directly (no fine-tuning) from standard STARBURST99 stellar population models \citep{leitherer1999} assuming a \cite{kroupa} IMF.}

\mybold{Feedback in FIRE-2 is implemented through a combination of several resolved channels: continuous stellar mass-loss, photoionisation, photoelectric heating, radiation pressure, and supernovae (Types Ia and II). Crucially, FIRE-2 mass and time resolution are sufficient to explicitly treat individual supernova explosions as discrete events rather than continuous energy injection. The code calculates the exact Sedov–Taylor solution for an energy-conserving spherical shock; if the resolved coupling radius between the star and gas is smaller than the cooling radius, the energy is injected thermally and the subsequent expansion is physically resolved. However, if the coupling radius is larger than the cooling radius, only the terminal momentum is deposited to account for the momentum-conserving phase that occurred at unresolved scales. This allows the simulation to resolve the multiphase structure of the interstellar medium by affecting baryons directly at small scales ($\sim 1-10$ pc) without ever turning off gas cooling.} Reionisation is treated via the uniform ultraviolet background of \cite{FG09}, leading to a fully ionised hydrogen intergalactic medium by z $\sim$ 6. These simulations reach high spatial resolution, with gravitational softening lengths on the order of 1-10 pc. \mybold{The simulations also feature exceptional mass resolution, with minimum baryonic particle masses ranging from 30 to $7\times10^{3}$ M$_{\odot}$, and dark matter particles spanning 200 to $4\times10^{4}$ M$_{\odot}$. Minimum time-steps are of the order of 100 yr, and the maximum time-step allowed for star particles is of 10$^{4}$ yr.}

The FIRE-2 simulations have been shown to reproduce a broad range of observed galaxy properties. As demonstrated in \cite{Hopkins_2018}, these include realistic star formation histories, gas distributions, metallicity profiles, morphologies, and rotation curves, as well as stellar mass scaling relations.

We use the FIRE-2 cosmological zoom-in simulations presented in \cite{Hopkins_2018} and in \cite{m10x}\mybold{, on top of high-resolution runs from \cite{m09res30}}. Haloes were again identified using AHF. Exploiting the higher resolution of the FIRE-2 simulations and aiming to improve the statistical robustness of our analysis, we did not restrict our selection to the primary halo in each simulation. Instead, for each simulation, we considered the ten haloes containing the largest number of particles. From these, we kept only those composed of at least 99\% high-resolution particles and containing a minimum of a hundred star particles. Applying these criteria results in a sample of 109 isolated galaxies.

\subsection{Previous Comparisons}
\label{sec:comparisons}

\mybold{Previous studies have compared the NIHAO and FIRE simulation suites to assess how their different numerical treatments of stellar feedback impact galaxy evolution. A primary focus of these comparisons is the cusp-to-core transformation in dark matter density profiles. Both NIHAO and FIRE-2 successfully reproduce the flattening of central dark matter distributions, transforming the cuspy profiles into cores \citep{Tollet, Lazar}. While both suites agree on the integrated mass profiles, the physical drivers differ; FIRE-2 is characterized by more bursty star formation that causes rapid fluctuations in the gravitational potential, whereas NIHAO relies on the large-scale energy injection of the thermal blast-wave model \citep{Tollet, NIHAO-RING}.}

\mybold{Recent comparisons using identical initial conditions reveal that while both codes produce similar final disc morphologies, they differ significantly in their circum-galactic medium properties \citep{NIHAO-RING}. NIHAO (GASOLINE) is characterized by generalised, long-term outflows that suppress inter-galactic medium accretion, while FIRE-2 (GIZMO) presents quasi-virialised hot gas halos formed by the interaction of inflows and feedback-driven outflows \citep{NIHAO-RING}.}

\section{Results}
\label{sec:results}
We defined the virial radius, $R_{\rm vir}$, as the radius enclosing a mean density equal to $\Delta_{\rm vir}$ times the critical density of the universe, $\rho_{\rm crit} = \frac{3H^{2}}{8\pi G}$, where $H$ is the Hubble parameter and $G$ is the gravitational constant. The value of $\Delta_{\rm vir}$ was calculated using the redshift-dependent prescription from \cite{virial_overdensity}, evaluated at $z=0$. The virial mass was then obtained by summing the masses of all particles within $R_{\rm vir}$, and the stellar mass was computed as the total mass of stellar particles within 0.1$R_{\rm vir}$.

To characterise the inner slope of the dark matter density profile, we computed a linear fit to the logarithmic density profile between 1\% and 2\% of $R_{\rm vir}$, as in \cite{dicintio_profile, dicintio_mstar_mhalo}. For this purpose, we adopted the approach from \cite{Lazar} to construct dark matter density profiles using 35 logarithmically spaced radial bins between 0.005 $R_{\rm vir}$ and $R_{\rm vir}$. This method was consistently applied to both NIHAO and FIRE-2 galaxies, differing from the binning used for the analysis of NIHAO simulations presented by \cite{Tollet}.

\begin{figure}[ht]
    \centering
    \includegraphics[width=\columnwidth]{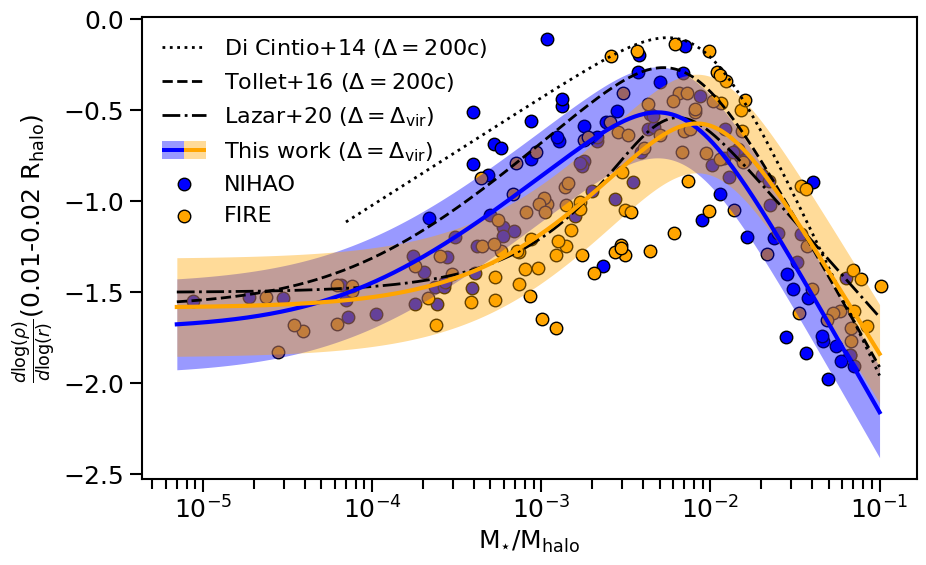}
    \caption{Inner slope of the dark matter density profile, measured between 1\% and 2\% of the virial radius, as a function of the stellar-to-halo mass ratio. Results are shown for galaxies from the NIHAO (blue) and FIRE-2 (orange) simulations. Solid lines represent fits following Eq.~\ref{eq:tollet}  with parameters from Table~\ref{tab:params-FIREandNIHAO} and the 1-$\sigma$ scatter around the fits is indicated with shadowed regions. The trends are compared to literature fits from \citet{dicintio_mstar_mhalo}, \citet{Tollet} and \citet{Lazar}.}
    \label{fig:main_figure_both_sims_together}
\end{figure}

In Fig.~\ref{fig:main_figure_both_sims_together}, we present the dark matter profile inner slope (measured between 1\% and 2\% of $R_{\rm vir}$) as a function of the stellar-to-halo mass ratio. We fitted each set of simulations independently using Eq.~\ref{eq:tollet}, and we provide the fitted parameters in Table~\ref{tab:params-FIREandNIHAO}. Our results align well with previous findings for FIRE-2 simulations \citep{Lazar} but notable discrepancies exist between our results for NIHAO simulations and those reported in \citep{Tollet}. This mismatch arises from the difference in the definition of virial overdensity used for the analysis, since they adopted the value $\Delta=\rm200c$\footnote{Throughout the text, the expression $\Delta=\rm200c$ denotes an overdensity of 200 times the critical density of the universe, and we use the letter c to explicitly differentiate it from an overdensity of 200 times the average matter density of the universe.}. The radius $R_{\rm 200c}$ is smaller than $R_{\rm vir}$, and the slope measured by taking it as a reference is inherently more cored than it would by using $R_{\rm vir}$. In Fig.~\ref{fig:main_figure_both_sims_together_R200} we show our results when matching this different overdensity definition. Even for FIRE-2 simulations, we note slight discrepancies with the fit from the literature. Some possible origin of these differences could be the algorithm used for centering galaxies or the specific selection of galaxies, as our sample includes more galaxies than previous works for both simulation suites. \mybold{\cite{Tollet} uses $\sim$70 of the NIHAO simulations from our more complete sample. \cite{Lazar} only considers the main halo of each FIRE-2 simulation, except for previously identified additional galaxies found in \cite{m10x}, contrasting with our previously described selection method. Also, they use some runs of the same initial conditions with different resolutions, while we only use the highest resolution available for each run.}

After homogenising the analysis across NIHAO and FIRE-2 galaxies, the relation for both simulation suites is in much closer agreement than it could be inferred from previous works, in which the trends derived in \cite{Tollet} and \cite{Lazar} are compared without taking into account the differences in the analysis. Whilst previous relations indicated that NIHAO galaxies are more efficient at developing shallow dark matter profiles, our results reveal that both simulation suites are capable of forming strong cores. The 1-$\sigma$ scatter around the relation is also similar: 0.25 for NIHAO and 0.27 for FIRE-2. However, some systematic differences between the two simulation suites remain: whereas we find the peak efficiency of core formation for NIHAO galaxies at $M_{\star}/M_{\rm vir}$ $\sim$ 0.005, FIRE-2 galaxies present their highest count of cored dark matter profiles at $M_{\star}/M_{\rm vir}$ $\sim$ 0.008.

\begin{table}[ht]
\caption{Fitted parameters for each set of simulations using Eq.~\ref{eq:tollet} and the overdensity $\Delta =\Delta_{\rm vir}$.}
\label{tab:params-FIREandNIHAO}
\begin{adjustbox}{max width=\columnwidth}
\begin{tabular}{ c c c c c c c }
\hline
 & n     & n$_{1}$ & x$_{1}$                 & x$_{0}$                 & $\beta$ & $\gamma$ \\ \hline
NIHAO       & -0.87 & 6.86    & 8.94 $\times$ 10$^{-5}$ & 1.61 $\times$ 10$^{-2}$ & 0.78    & 1.63     \\ \hline
FIRE-2        & -1.01 & 3.80    & 8.54 $\times$ 10$^{-4}$ & 2.83 $\times$ 10$^{-2}$ & 1.21    & 1.52     \\ \hline
\end{tabular}%
\end{adjustbox}
\end{table}

\subsection{Burstiness}
\label{sec:results-burst}

To quantify burstiness in a galaxy’s SFH, we defined a starburst phase using the following criterion:

\begin{equation}
\label{eq:burst}
    \left\langle {\rm SFR} (50 \; \rm Myr)\right\rangle > 1.5 \left\langle {\rm SFR} (500 \; \rm Myr)\right\rangle
\end{equation}

This is conceptually similar to the approach in \cite{FIRE_burst}, although we used longer time windows (50 and 500 Myr, compared to their 10 and 200 Myr). Using this definition, we calculated a metric: the \emph{bursty mass fraction} \mybold{($f_{\rm M, burst}$), defined as the fraction of a galaxy's stellar mass formed during starburst phases.}

\begin{figure}[ht]
    \centering
    \includegraphics[width=\columnwidth]{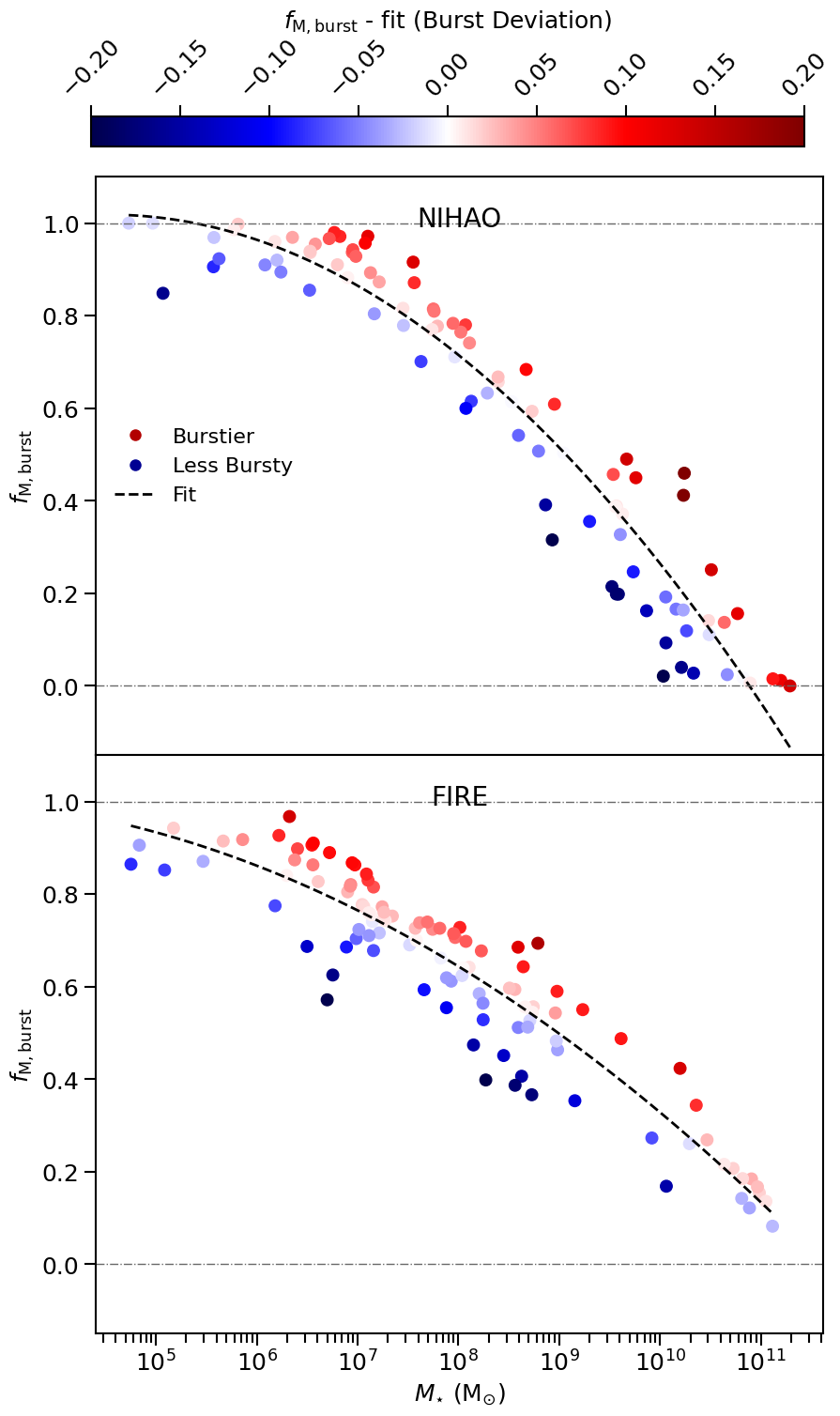}
    \caption{Bursty mass fractions as a function of stellar mass for galaxies in NIHAO (top panel) and FIRE-2 (bottom panel) simulation suites. Black dashed lines show the result of fitting the bursty mass fraction to the stellar mass with a second degree polynomial. Circles are coloured by their absolute deviation from the fit, which we define as \emph{burst deviation}.}
    \label{fig:burstiness_stellarmass_2panels}
\end{figure}

Fig.~\ref{fig:burstiness_stellarmass_2panels} shows how this quantity varies with stellar mass. \mybold{Fig. \ref{fig:sfr_3_panels} presents the SFH of five NIHAO galaxies of different stellar masses, illustrating visually the dependence between burstiness and stellar mass.} Compared to the results in \cite{FIRE_burst}, our criterion produces a more gradually decreasing trend with stellar mass for FIRE-2 galaxies (they did not analyse NIHAO galaxies), which can be well approximated by a second degree polynomial. To characterise individual deviations from this trend, we defined the \emph{burst deviation} as the difference between a galaxy’s bursty mass fraction and the value predicted from the polynomial fit at its stellar mass. This deviation quantifies how much more or less bursty a galaxy is than expected for its stellar mass.

\begin{table}[ht]
\caption{Fitted parameters for galaxies with positive and negative burst deviation in each set of simulations using Eq.~\ref{eq:tollet} and the overdensity $\Delta =\Delta_{\rm vir}$.}
\label{tab:params-FIREandNIHAO-burst}
\begin{adjustbox}{max width=\columnwidth, height = 1.65cm}
\begin{tabular}{ c c c c c }
\hline
 & NIHAO ($>$0) & NIHAO ($<$0) & FIRE-2 ($>$0) & FIRE-2 ($<$0) \\ \hline
n & -3.11 & -1.33 & -0.73 & -0.61 \\ \hline
n$_{1}$ & 119 & 1.65 & 6.52 & 7.28 \\ \hline
x$_{1}$ & 4.71 $\times$ 10$^{-9}$ & 1.85 $\times$ 10$^{-3}$ & 1.47 $\times$ 10$^{-3}$ & 0.93 \\ \hline
x$_{0}$ & 1.13 & 2.63 $\times$ 10$^{-2}$ & 1.69 $\times$ 10$^{-2}$ & 1.26 $\times$ 10$^{-2}$ \\ \hline
$\beta$ & 0.83 & 2.27 & 1.78 & 348 \\ \hline
$\gamma$ & 1.26 & 1.60 & 1.52 & 1.16 \\ \hline
\end{tabular}
\end{adjustbox}
\end{table}

\begin{figure}[ht]
    \centering
    \includegraphics[width=1.025\columnwidth]{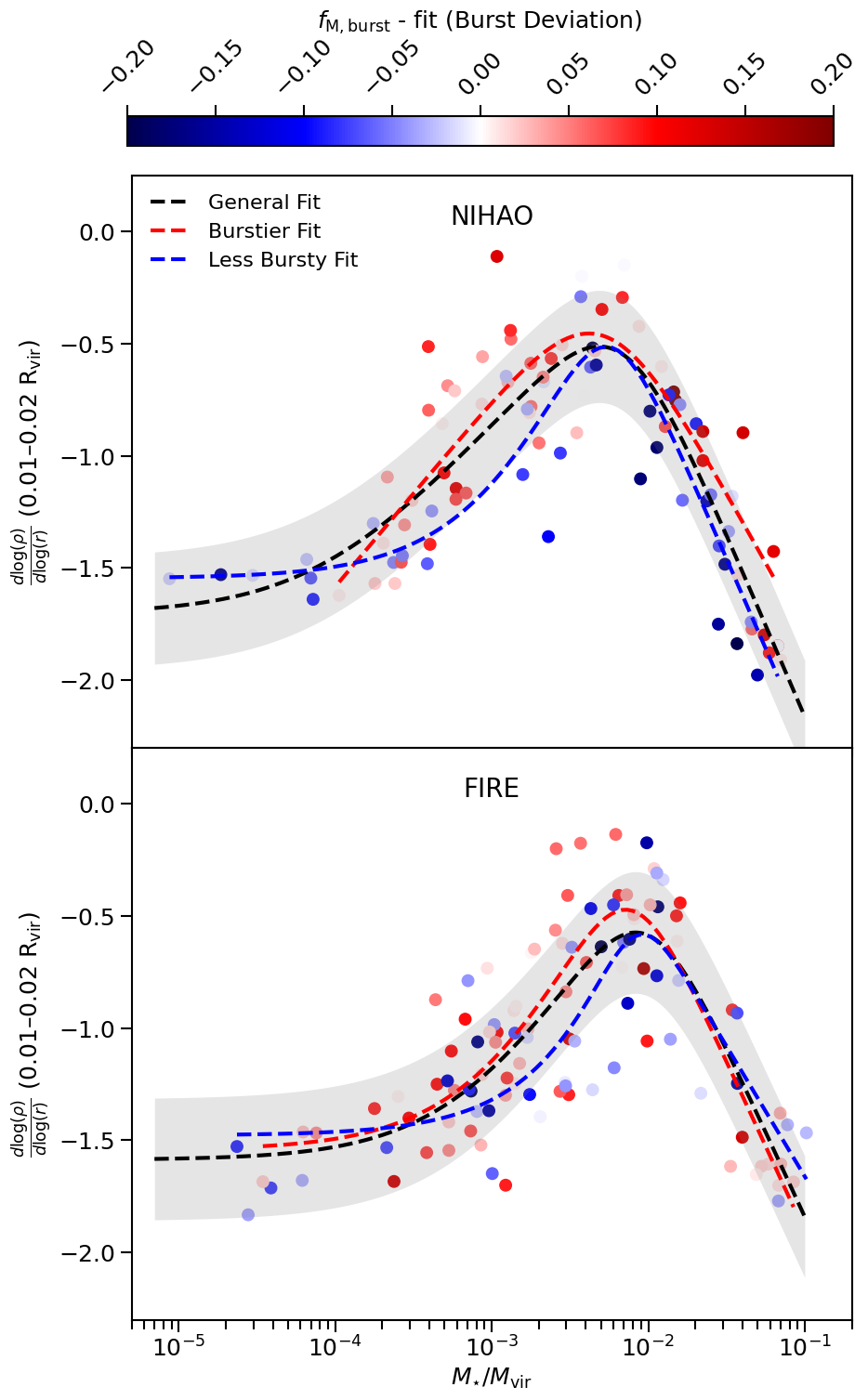}
    \caption{Inner slope of the dark matter density profile, measured between 1\% and 2\% of the virial radius, as a function of the stellar-to-halo mass ratio. The top panel shows results for NIHAO galaxies and the bottom panel shows results for FIRE-2 galaxies. Circles for each galaxy are colour coded by their \emph{burst deviation} (the difference between actual and expected bursty mass fraction at a given stellar mass). Separate fits are shown for galaxies with above- (red dashed lines) and below- (blue dashed lines) average burstiness. Black dashed lines show the fits to the full sample of galaxies from each suite and gray bands represent the 1-$\sigma$ scatter around the fit. All fits follow Eq.~\ref{eq:tollet} with parameters shown in Table~\ref{tab:params-FIREandNIHAO-burst}.}
    \label{fig:main_figure_color_bursty}
\end{figure}

In Fig.~\ref{fig:main_figure_color_bursty}, we plot the inner slope of the dark matter density profile as a function of the stellar-to-halo mass ratio for both NIHAO and FIRE-2 galaxies. This time, each galaxy is coloured by its burst deviation. We fitted separate trends using Eq.~\ref{eq:tollet} for galaxies with positive (more bursty than expected) and negative (less bursty than expected), yielding the parameters in Table~\ref{tab:params-FIREandNIHAO-burst}.

For NIHAO galaxies, the trend is clear: galaxies that are burstier than the average for their stellar mass are more likely to exhibit cored dark matter profiles. The trend is similar in the FIRE-2 sample for mass ratios at and just below the peak in core formation ($M_{\star}/M_{\rm vir}$ $\sim$ 0.005-0.008) but disappears for $M_{\star}/M_{\rm vir}$ > 0.02, where we find a gap in the galaxies from the sample.

\subsection{\mybold{Post-to-pre reionisation stellar mass ratio}}
\label{sec:results-mpostpre}
In addition to burstiness, we explore how the \mybold{temporal concentration} of star formation relative to reionisation affects the inner dark matter slope. We define the accumulated stellar mass before and after reionisation as:

\begin{equation}
\label{eq:mpre}
    M_{\star, \rm pre} = \int^{t_{\rm reion}}_{t=0} {\rm SFR} dt
\end{equation}
\begin{equation}
\label{eq:mpost}
    M_{\star, \rm post} = \int^{t_{\rm today}}_{t=t_{\rm reion}} {\rm SFR} dt
\end{equation}

\begin{figure}[ht]
    
    \centering
    \includegraphics[width=\columnwidth]{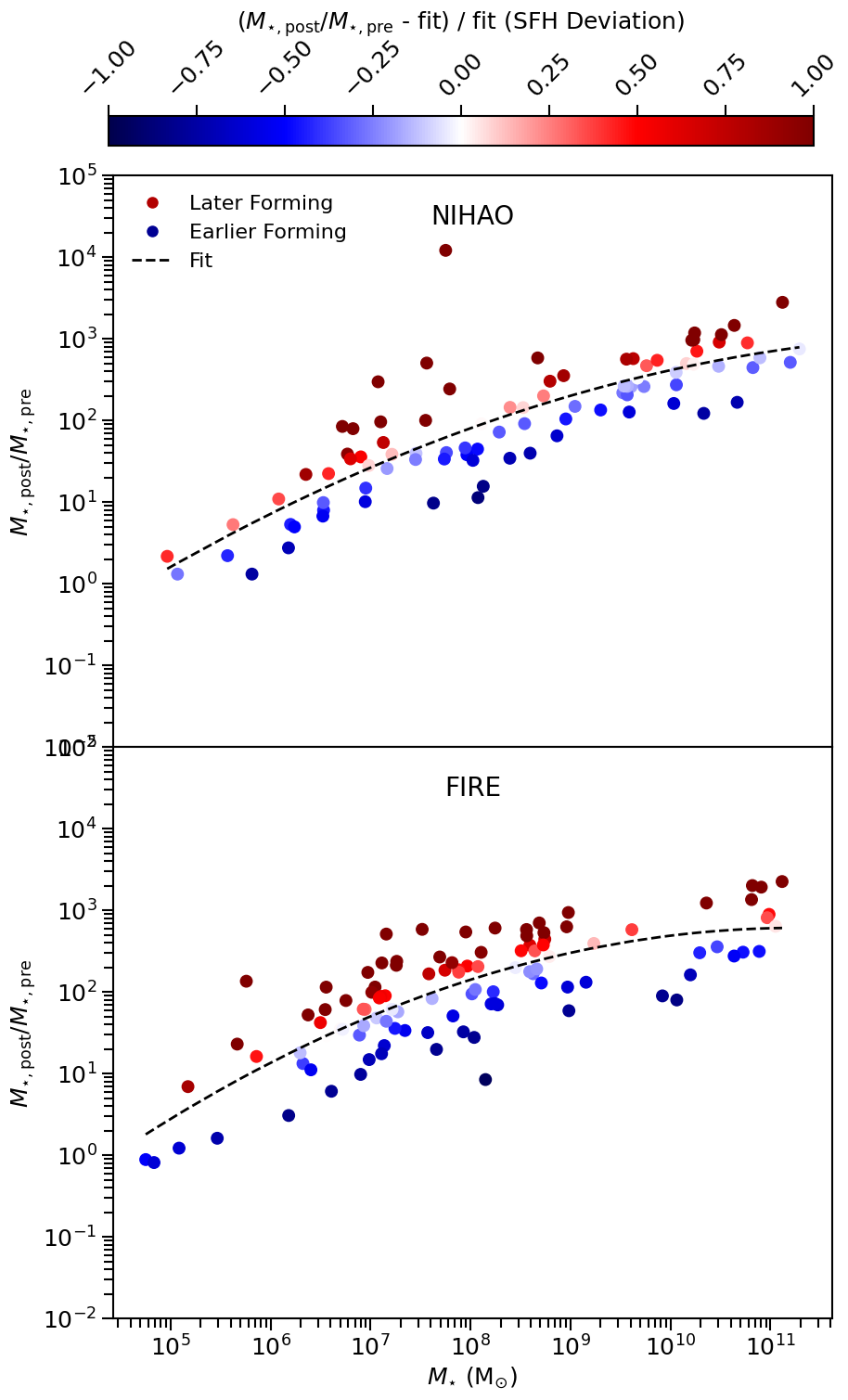}
    \caption{Ratio of post-to-pre reionisation stellar mass ($M_{\star, \rm post}/M_{\star, \rm pre}$) as a function of total stellar mass for galaxies in NIHAO (top panel) and FIRE-2 (bottom panel) simulation suites. Black dashed lines show second degree polynomial fits. Circles are coloured by their relative deviation from the fit, which we define as \emph{SFH deviation}.}
    \label{fig:mpostpre_stellarmass_2panels} 
\end{figure}

\begin{figure}[ht]
    
    \centering
    \includegraphics[width=\columnwidth]{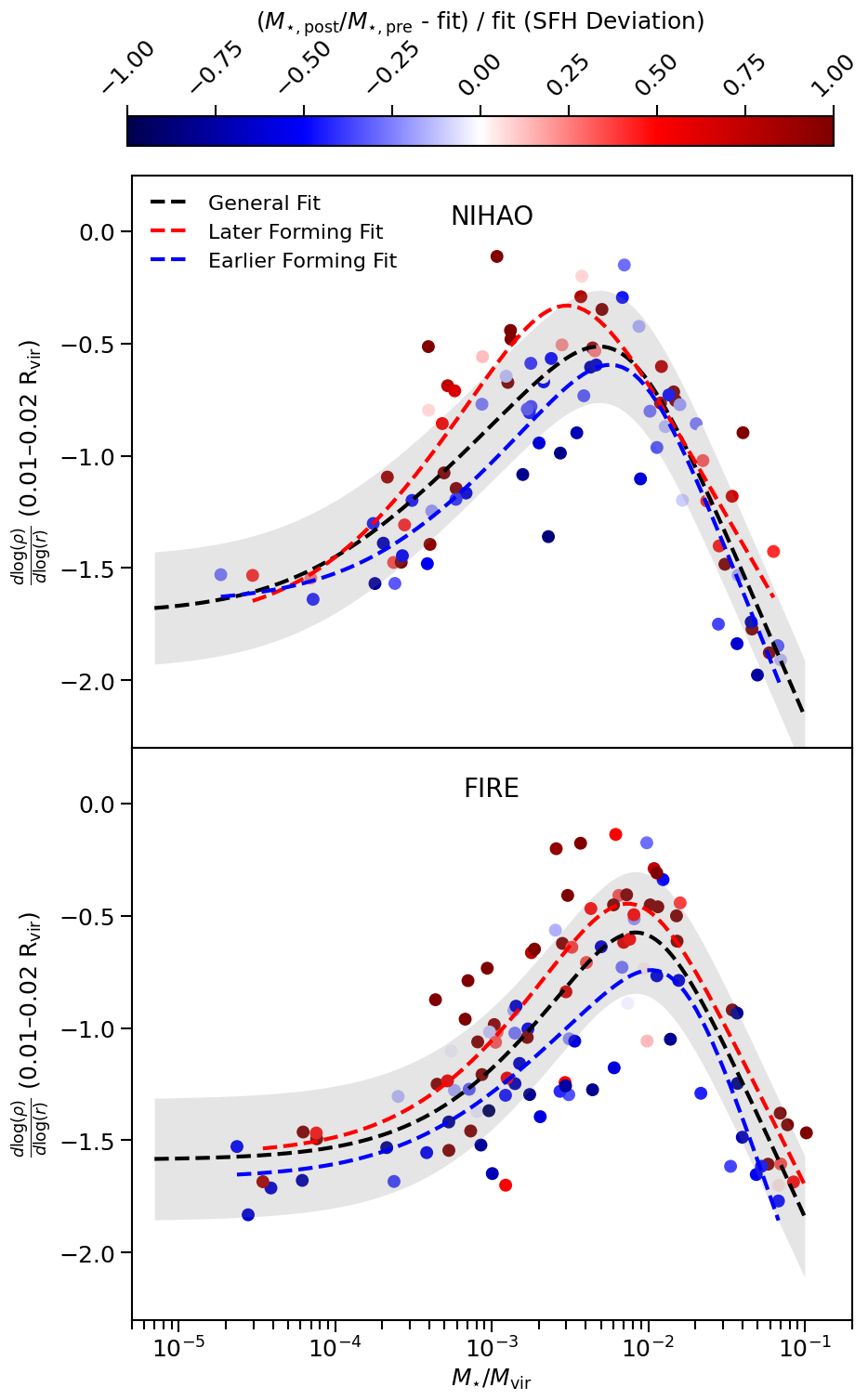}
    \caption{Inner slope of the dark matter density profile, measured between 1\% and 2\% of the virial radius, as a function of the stellar-to-halo mass ratio. The top panel shows results for NIHAO galaxies and the bottom panel shows results for FIRE-2 galaxies. Circles for each galaxy are colour coded by their \emph{SFH deviation} (the relative difference between actual and expected post-to-pre reionisation stellar mass ratio at a given stellar mass). Separate fits are shown for galaxies with more extended (red dashed lines) and less extended (blue dashed lines) SFHs. Black dashed lines show the fits to the full sample of galaxies from each suite and gray bands represent the 1-$\sigma$ scatter around the fit. All fits follow Eq.~\ref{eq:tollet} with parameters shown in Table~\ref{tab:params-FIREandNIHAO-mpostpre}.}
    \label{fig:main_figure_color_Mpostpre}
\end{figure}

These follow the definitions in \cite{Muni_2024}, with $z_{\rm reion}$ = 6.5 ($t_{\rm reion} = 0.84$ Gyr). This is motivated by the \cite{FG09} ultraviolet background in EDGE/FIRE-2, though the reionisation timescale in NIHAO is comparable despite its different background model.

The ratio $M_{\star, \rm post}$ / $M_{\star, \rm pre}$ \mybold{acts as a proxy for the temporal concentration} of the SFH\footnote{We exclude a small subsample of galaxies lacking star formation prior to z = 6.5, resulting in a final sample of 90 NIHAO and 108 FIRE-2 galaxies.} and increases with stellar mass (Fig.~\ref{fig:mpostpre_stellarmass_2panels}). As expected, more massive galaxies are more efficient at converting baryons into stars, \mybold{and less affected by feedback due to early star formation episodes, leading to higher $M_{\star, \rm post}$ / $M_{\star, \rm pre}$ values. This trend is visualized in Fig.~\ref{fig:mpostpre_mhalo_vs_mstar}}. Note that both the NIHAO and FIRE-2 galaxies in our sample are isolated and not subject to quenching due to environmental factors. To quantify individual deviations from this trend, we defined the \emph{SFH deviation} as the fractional difference between a galaxy’s $M_{\star, \rm post}$ / $M_{\star, \rm pre}$ ratio and the value predicted by a second degree polynomial fit to the stellar mass.

We analyse the trends in inner dark matter slope versus stellar-to-halo mass ratio as a function of \emph{SFH deviation} in Fig.~\ref{fig:main_figure_color_Mpostpre}, with fit parameters for Eq.~\ref{eq:tollet} listed in Table~\ref{tab:params-FIREandNIHAO-mpostpre}. In both simulation suites, galaxies with extended SFHs (higher $M_{\star, \rm post}$ / $M_{\star, \rm pre}$ than expected) exhibit shallower dark matter profiles than those with more concentrated, early-time star formation.

\begin{table}[ht]
\caption{Fitted parameters for galaxies with positive and negative SFH deviation in each set of simulations using Eq.~\ref{eq:tollet} and the overdensity $\Delta =\Delta_{\rm vir}$.}
\label{tab:params-FIREandNIHAO-mpostpre}
\begin{adjustbox}{max width=\columnwidth}
\begin{tabular}{ c c c c c }
\hline
 & NIHAO ($>$0) & NIHAO ($<$0) & FIRE-2 ($>$0) & FIRE-2 ($<$0) \\ \hline
n & -1.00 & -0.83 & -0.92 & -0.99 \\ \hline
n$_{1}$ & 5.73 & 6.63 & 4.37 & 4.73 \\ \hline
x$_{1}$ & 1.34 $\times$ 10$^{-4}$ & 2.44 $\times$ 10$^{-4}$ & 6.43 $\times$ 10$^{-4}$ & 4.88 $\times$ 10$^{-4}$\\ \hline
x$_{0}$ & 1.90 $\times$ 10$^{-2}$ & 1.48 $\times$ 10$^{-2}$ & 2.84 $\times$ 10$^{-2}$ & 2.61 $\times$ 10$^{-2}$ \\ \hline
$\beta$ & 1.26 & 0.89 & 1.26 & 0.79 \\ \hline
$\gamma$ & 1.21 & 1.76 & 1.42 & 2.06 \\ \hline
\end{tabular}
\end{adjustbox}
\end{table}

\cite{Muni_2024} performed additional analyses of how the $M_{\star, \rm post}$ / $M_{\star, \rm pre}$ ratio influences dark matter profile shapes, employing alternative coreness indicators rather than the 1–2\% $R_{\rm vir}$ slope used in this work. We reproduce their tests and present the corresponding results in Appendix~\ref{appendixMUNI}.

\subsection{Second order correction to the inner slope fit}
\label{sec:secondorderfit}

\begin{figure}[H]
    \centering
    \includegraphics[width=\columnwidth]{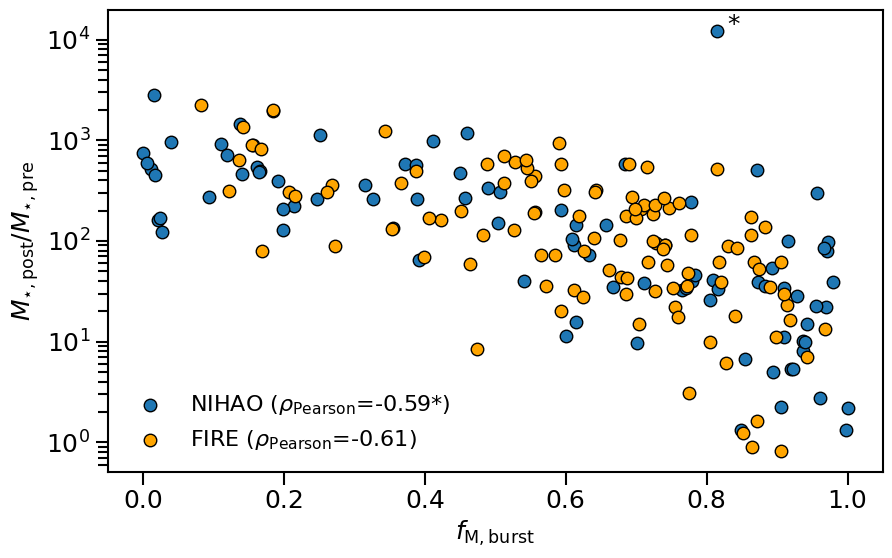}
    \caption{Relation between $M_{\star, \rm post}$ / $M_{\star, \rm pre}$ and the bursty mass fraction for NIHAO and FIRE-2 galaxies. The Pearson correlation coefficient is indicated in the legend. One outlier NIHAO galaxy, marked with an asterisk, is removed from the calculation.}
    \label{fig:burtiness_mpostpre_correation}
\end{figure}

We have demonstrated that both burstiness and the \mybold{post-to-pre reionisation stellar mass ratio} play a role, separate from the stellar-to-halo mass ratio, in shaping the inner region of the dark matter density profile of halos. We now aim to use these features to obtain a correction term over the fitting by Eq.~\ref{eq:tollet}. We decided to use only one of the features at a time, since we find them to be correlated, as we show in Fig.~\ref{fig:burtiness_mpostpre_correation}. In that figure we also notice an outlier galaxy with a very high $M_{\star, \rm post}$ / $M_{\star, \rm pre}$ value, which can also be observed in Fig.~\ref{fig:mpostpre_stellarmass_2panels}. While this galaxy does not introduce any notable effect in the previous analysis, we decide to remove it for the fitting performed in this section.

\begin{figure}[htb]
    \centering
\includegraphics[width=\columnwidth]{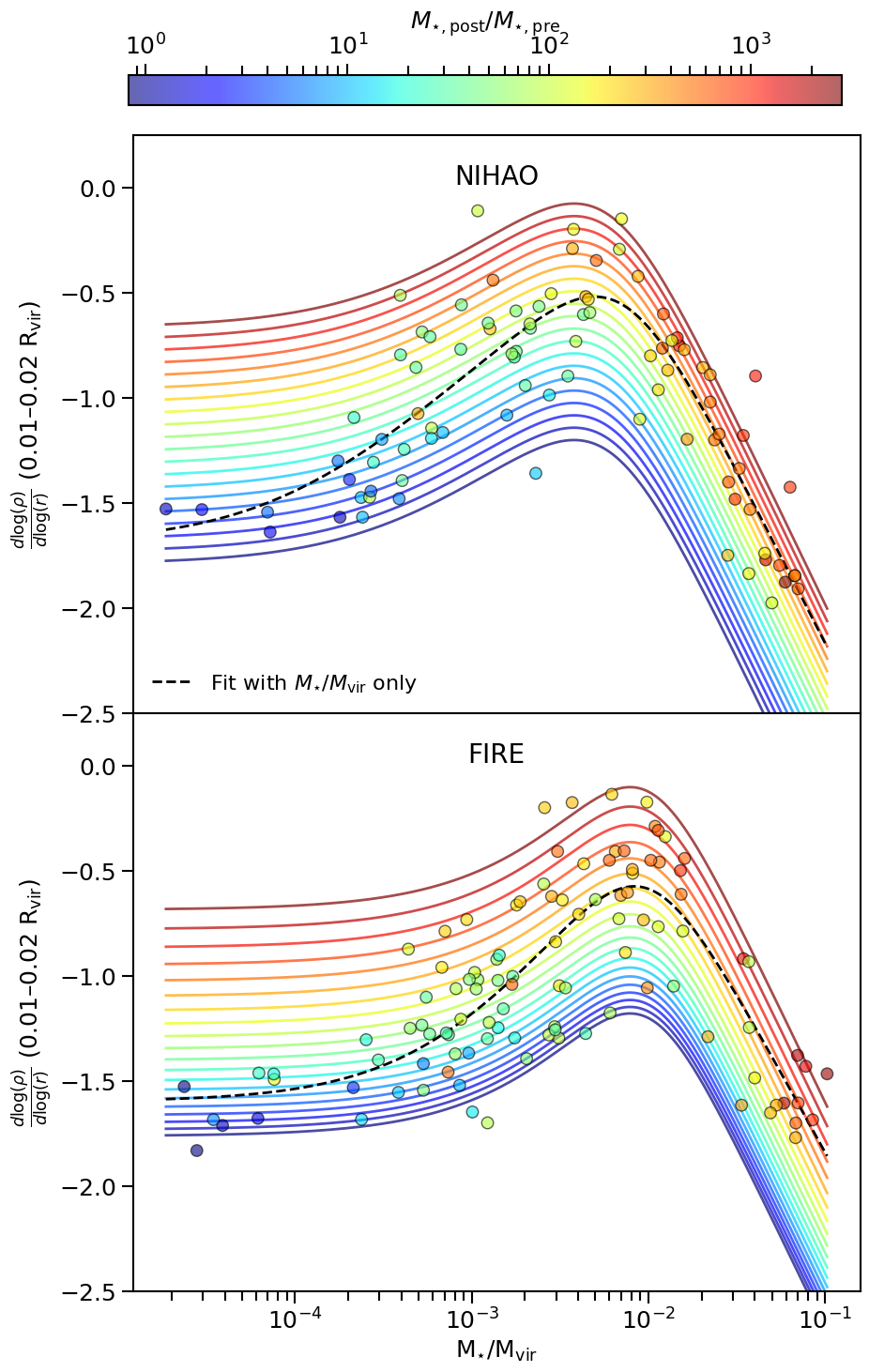}
    \caption{True values (circles) and predictions (lines) for the inner slope of the dark matter density profile as a function of the stellar-to-halo mass ratio for different values of $M_{\star, \rm post}$ / $M_{\star, \rm pre}$ via the fit by Eq.~\ref{eq:tollet-corrected} with the parameters shown in Table~\ref{tab:params-FIREandNIHAO-corrected}. Black dashed lines indicate fits using only the stellar-to-halo mass ratio, described by the parameters in Table \ref{tab:params-FIREandNIHAO} for Eq. \ref{eq:tollet}. The top panel shows the results for the NIHAO suite and the bottom panel corresponds to the galaxies from the FIRE-2 dataset.}
    \label{fig:corrected-fit}
\end{figure}

We report a two variable fitting formula for the inner slope described by Eq.~\ref{eq:tollet-corrected}, where x = $M_{\star}/M_{\rm vir}$ and y = $M_{\star, \rm post}$ / $M_{\star, \rm pre}$. The inclusion of the additional correction lowers the mean squared error in the prediction of the inner slope from 0.073 to 0.055 for FIRE-2 galaxies, and from 0.063 to 0.048 for NIHAO galaxies. We show the fitted parameters for each simulation suite in Table~\ref{tab:params-FIREandNIHAO-corrected}. In Fig.~\ref{fig:corrected-fit} we visualise the fit by plotting the trend of the inner slope with the stellar-to-halo mass ratio for different values of $M_{\star, \rm post}$ / $M_{\star, \rm pre}$. In the figure we can observe how, for the same ratio of stellar mass to halo mass, galaxies are more likely to present shallower density profiles when they have undergone extended SFHs (high values of $M_{\star, \rm post}$ / $M_{\star, \rm pre}$).

\begin{equation}
\label{eq:tollet-corrected}
\begin{split}
\left.\frac{d\log{\rho}}{d\log{r}}\right|_{r = 0.015 R_{\rm halo}} (x) &= n - \log_{10} \left[ n_{1} \left( 1+\frac{x}{x_{1}}\right)^{-\beta} + \left( \frac{x}{x_{0}}\right)^\gamma \right] \\
&+ n_{2} y^{\alpha}
\end{split}
\end{equation}

\begin{table}[ht]
\caption{Fitted parameters for each set of simulations using Eq.~\ref{eq:tollet-corrected} and the overdensity $\Delta =\Delta_{\rm vir}$.}
\label{tab:params-FIREandNIHAO-corrected}
\begin{adjustbox}{max width=\columnwidth}
\begin{tabular}{ c c c c c c c c c }
\hline
 & n     & n$_{1}$& n$_{2}$ & x$_{1}$                 & x$_{0}$                 & $\beta$ & $\gamma$& $\alpha$ \\ \hline
NIHAO       & -66.9& 34.7 & 66.7    & 3.57 $\times$ 10$^{-4}$ & 2.34 $\times$ 10$^{-3}$ & 0.68    & 1.75  &  2.06 $\times$ 10$^{-3}$  \\ \hline
FIRE-2        & -2.10& 1.44 & 0.51    & 3.16 $\times$ 10$^{-3}$ & 2.43 $\times$ 10$^{-2}$ & 1.45    & 1.75  &  0.14 \\ \hline
\end{tabular}%
\end{adjustbox}
\end{table}

\section{Conclusions}
\label{sec:conclusions}

In this work, we revisited the dependence of the inner slope of dark matter density profiles on the stellar-to-halo mass ratio using homogeneously analysed samples from the NIHAO and FIRE-2 simulations. Our results bring the two suites into closer agreement, showing that careful methodological consistency is essential when comparing simulations and interpreting apparent differences across studies. Whereas earlier comparisons between the analyses of \cite{Tollet} and \cite{Lazar} suggested that NIHAO galaxies were intrinsically more prone to developing cored dark matter profiles than those in FIRE-2, our homogeneous reanalysis shows that both simulation suites are capable of producing galaxies with similarly cored inner density structures. However, the stellar-to-halo mass ratio at which core formation is most efficient is slightly lower for NIHAO galaxies than for the FIRE-2 suite.

We quantified star formation burstiness by adapting the burst criterion introduced in \cite{FIRE_burst}, deriving burst mass fractions that trace how concentrated in time star formation is within a galaxy. We find that burstiness decreases systematically with increasing stellar mass in the range M$_{\star} \sim 10^{5} - 10^{11}$ M$_{\odot}$.

To characterise the temporal \mybold{temporal concentration} of SFHs, we employed the $M_{\star, \rm post} / M_{\star, \rm pre}$ ratio introduced by \cite{Muni_2024}. In both NIHAO and FIRE-2 galaxies, higher stellar mass systems exhibit more extended SFHs.

By fitting the dependence of burstiness and SFH \mybold{post-to-pre reionisation stellar mass ratio} on stellar mass, we identified residuals that highlight galaxies which are more or less bursty, or have longer or shorter SFHs, than the average for their mass. These deviations act as secondary predictors of dark matter core formation, beyond the primary role of the stellar-to-halo mass ratio. Specifically, we find that galaxies which are burstier than average develop shallower inner dark matter slopes. Similarly, those with more extended SFHs are more efficient at producing cores. These SFH features explain the existing scatter in the relation between the inner slope and the stellar-to-halo mass ratio in both simulation suites.

We also derived a new fitting formula for the inner slope that includes the contribution of the SFH \mybold{temporal concentration} via the  $M_{\star, \rm post} / M_{\star, \rm pre}$ ratio (Eq.~\ref{eq:tollet-corrected}). The addition of a simple term and the joint fitting to data from each simulation revealed higher accuracy than using only the stellar-to-halo mass ratio as a fitting variable, lowering the mean squared error from 0.073 to 0.055 for FIRE-2 galaxies and from 0.063 to 0.048 for NIHAO simulations.

Our analysis suggests that burstiness and SFH \mybold{temporal concentration} capture important aspects of the baryonic processes that shape dark matter profiles. However, additional factors, such as the relative strength and timing of starbursts or their spatial distribution within galaxies, are likely to play a significant role. Exploring these features will be the focus of future work.

The predicted relationship between core formation and SFH at fixed stellar mass has significant observational implications. Advances in resolved stellar population studies now enable accurate SFH measurements for low mass galaxies well beyond the virial radius of the Milky Way \citep[e.g.][]{weisz_bullock, cohen_bullock}. We predict that galaxies with stellar masses of M$_{\star}\sim10^{8}$M$_{\odot}$, situated near the peak of core formation efficiency (M$_{\star}$/M$_{\rm vir}\sim0.005$), should display structural differences driven by their assembly age. Specifically, systems with stellar age distributions biased towards early times are expected to retain cuspier profiles than similar mass galaxies with extended, late time star formation. Observing large cores in objects dominated by early star formation would contradict this prediction, suggesting that feedback mechanisms alone are insufficient and pointing toward alternative scenarios such as Self-Interacting Dark Matter (SIDM).

\begin{acknowledgements}
 JSA thanks the Spanish Ministry of Economy and Competitiveness (MINECO) for support through a grant P/301404 from the Severo Ochoa project CEX2019-000920-S and the Fostering Grads organisation for sponsoring the ECUSA program, which made this international collaboration possible. JSB is supported by NSF grant AST-2408246. CB is supported by the Spanish Ministry of Science and Innovation (MICIU/FEDER) through research grant PID2021-122603NBC22. ADC is supported by the Agencia Estatal de Investigación, under the 2023 call for Ayudas para Incentivar la Consolidación Investigadora, grant number CNS2023-144669, project "TINY". The authors wish to acknowledge the contribution of the IAC High-Performance Computing support team and hardware facilities to the results of this research. The freely available software pynbody \citep{pynbody} has been used for part of this analysis. We thank the FIRE collaboration for making some simulations available for public use \citep{fire_public_1, fire_public_2} and for allowing the use of additional ones for this work.
\end{acknowledgements}

\bibliographystyle{aa}
\bibliography{bib.bib}

\appendix

\section{Analysing the inner slope with \texorpdfstring{$\Delta$}{Delta}~=~200c}
\label{appendixR200}

For this work, we adopted the virial overdensity $\Delta_{\rm vir}$ defined in \cite{virial_overdensity} to analyse the FIRE-2 and NIHAO simulations homogeneously. Nonetheless, several studies \citep[e.g.][]{dicintio_mstar_mhalo, Tollet} use a fixed overdensity of $\Delta = \rm 200c$ to investigate the relation between the inner slope of the dark matter density profile and the stellar-to-halo mass ratio. To facilitate a direct comparison to these works, we repeat our analysis using $\Delta = \rm 200c$. For this purpose, we construct new density profiles analogous to those presented in Fig. 2 of \cite{Tollet}, employing 56 logarithmically spaced bins spanning the range $0.01$–$1~R_{\rm 200c}$.

\begin{figure}[ht]
    \centering
    \includegraphics[width=\columnwidth]{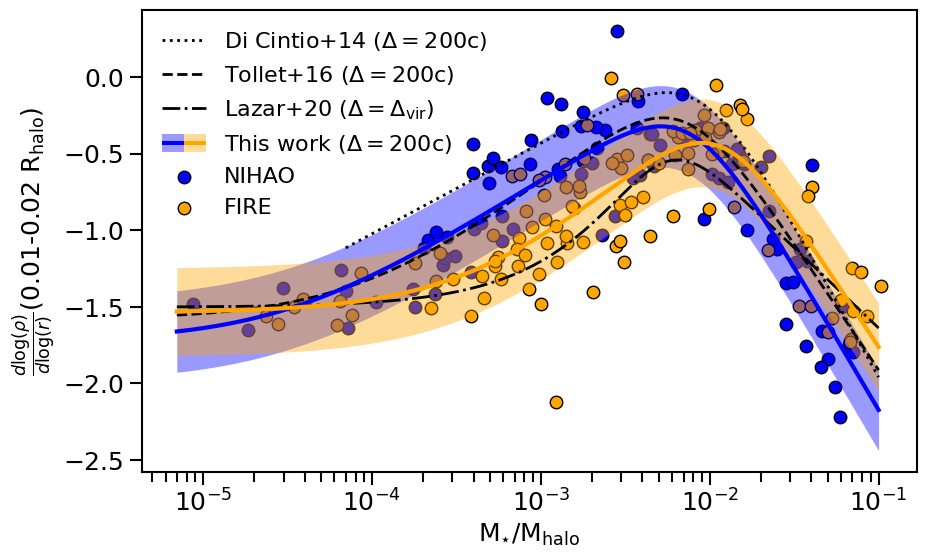}
    \caption{Inner slope of the dark matter density profile, measured between 1\% and 2\% of R$_{\rm 200c}$, as a function of the stellar-to-halo mass ratio. Results are shown for galaxies from the NIHAO (blue) and FIRE-2 (orange) simulations. Solid lines represent fits following Eq.~\ref{eq:tollet} and the 1-$\sigma$ scatter around the fits is indicated with shadowed regions. The trends are compared to literature fits from \citet{dicintio_mstar_mhalo}, \citet{Tollet} and \citet{Lazar}.}
    \label{fig:main_figure_both_sims_together_R200}
\end{figure}

We again apply Eq.~\ref{eq:tollet} to fit the inner slope of the dark matter density profile, measured between 1–2\% $R_{\rm 200c}$, as a function of the stellar-to-halo mass ratio $M_{\star}$/$M_{\rm 200c}$ (see fitted parameters in Table~\ref{tab:params-FIREandNIHAO-200}). The resulting relation is shown in Fig.~\ref{fig:main_figure_both_sims_together_R200}. Using the overdensity definition adopted by \cite{Tollet} produces a relation more consistent with their results, and galaxies with shallower profiles display even flatter inner slopes due to the more internal measurement radius. The scatter in the relation also increases, reaching 0.27 for NIHAO and 0.29 for FIRE-2. In addition, the stellar-to-halo mass ratio at which the profiles are maximally shallow shifts to higher values: 0.005 for NIHAO and 0.009 for FIRE-2. We note that the NIHAO peak value has increased slightly relative to our previous analysis, even though it rounds to the same value at one significant digit.

\begin{table}[ht]
\caption{Fitted parameters for each set of simulations using Eq.~\ref{eq:tollet} and the overdensity $\Delta~=~200$c.}
\label{tab:params-FIREandNIHAO-200}
\begin{adjustbox}{max width=\columnwidth}
\begin{tabular}{ c c c c c c c }
\hline
 & n     & n$_{1}$ & x$_{1}$                 & x$_{0}$                 & $\beta$ & $\gamma$ \\ \hline
NIHAO       & -1.09 & 4.31    & 3.24 $\times$ 10$^{-5}$ & 2.63 $\times$ 10$^{-2}$ & 0.70    & 1.88     \\ \hline
FIRE-2        & -0.85 & 4.86    & 4.43 $\times$ 10$^{-4}$ & 2.90 $\times$ 10$^{-2}$ & 0.98    & 1.70     \\ \hline
\end{tabular}
\end{adjustbox}
\end{table}

\section{Comparing results between EDGE and FIRE-2}
\label{appendixMUNI}

\begin{figure}[t]
    \centering
    \includegraphics[width=\columnwidth]{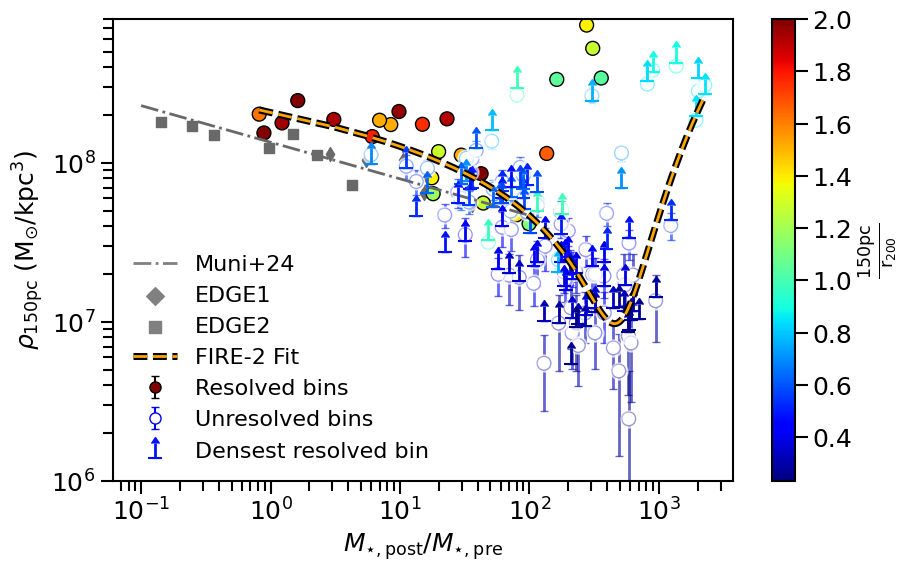}
    \caption{Dark matter density at 150 pc versus the ratio of stellar mass formed after and before reionisation. Gray symbols indicate \cite{Muni_2024} results for EDGE galaxies, and the dash-dotted line shows their linear fit. Filled (empty) circles indicate the inner densities of FIRE-2 galaxies that are (un)resolved at 150 pc, with errorbars representing Poisson uncertainties associated to the number of dark matter particles used to calculate the inner density. Markers are colored based on the ratio between 150 pc and r$_{200}$, \mybold{a resolution limit marked by the radius enclosing 200 dark matter particles}. For unresolved points, horizontal lines provide a lower limit to the inner density by taking the densest resolved bin in the density profile. The orange dashed line represents the fit to the inner densities, including unresolved bins.
    }
    \label{fig:rho_vs_Mpostpre}
\end{figure}

\cite{Muni_2024} analysed core formation in EDGE1 and EDGE2 simulations \citep{edge1, edge2} by measuring the dark matter density at 150 pc from galaxy centers. They reported a tight, decreasing linear relation between the inner dark matter density and the ratio $M_{\star, \rm post} / M_{\star, \rm pre}$. We replicate this analysis using FIRE-2 galaxies to calculate their density within a spherical shell ranging from 125 to 175 pc, with results shown in Fig.~\ref{fig:rho_vs_Mpostpre}. NIHAO galaxies are excluded due to insufficient spatial resolution. For some FIRE-2 galaxies, the 150 pc region lies within the $r_{200}$\footnote{The radius that encloses 200 dark matter particles.} resolution limit; however, we check that extrapolating the resolved density profile confirms that the inner densities do not significantly deviate from expected values. Additionally, we plot the density at the innermost resolved bin, providing a lower limit for the density at 150 pc.

Within the range of $M_{\star, \rm post} / M_{\star, \rm pre}$ values probed by EDGE simulations, we also recover a decreasing linear relation between inner density and $M_{\star, \rm post} / M_{\star, \rm pre}$. The trend is steeper for FIRE-2 galaxies, which also exhibit higher scatter. At higher $M_{\star, \rm post} / M_{\star, \rm pre}$ ratios, however, FIRE-2 galaxies show increased inner densities. We attribute this trend change to the broader stellar mass range in our sample. The relation for FIRE-2 galaxies is fitted using Eq.~\ref{eq:dicintio-inverse} with $x = M_{\star, \rm post} / M_{\star, \rm pre}$. The resulting fitted parameters are:  $x_{0} = 2.14 \times 10^{3}$, $n = - 6.22 \times 10^{-2}$, $\beta = 8.12 \times 10^{-2}$ and $\gamma = 1.98$. The resulting fitted relation is qualitatively similar to the classical relation between the dark matter density inner slope and the stellar-to-halo mass ratio (see e.g. Fig.~\ref{fig:main_figure_both_sims_together} and Eq.~\ref{eq:dicintio}).

\begin{equation}
\label{eq:dicintio-inverse}
\frac{\rho_{\rm 150pc}(x)}{10^{9} M_{\odot} kpc^{-3}} = n + \log_{10} \left[ \left( \frac{x}{x_{0}}\right)^{-\beta} + \left( \frac{x}{x_{0}}\right)^\gamma \right]
\end{equation}

\begin{figure}[t]
    \centering
    \includegraphics[width=\columnwidth]{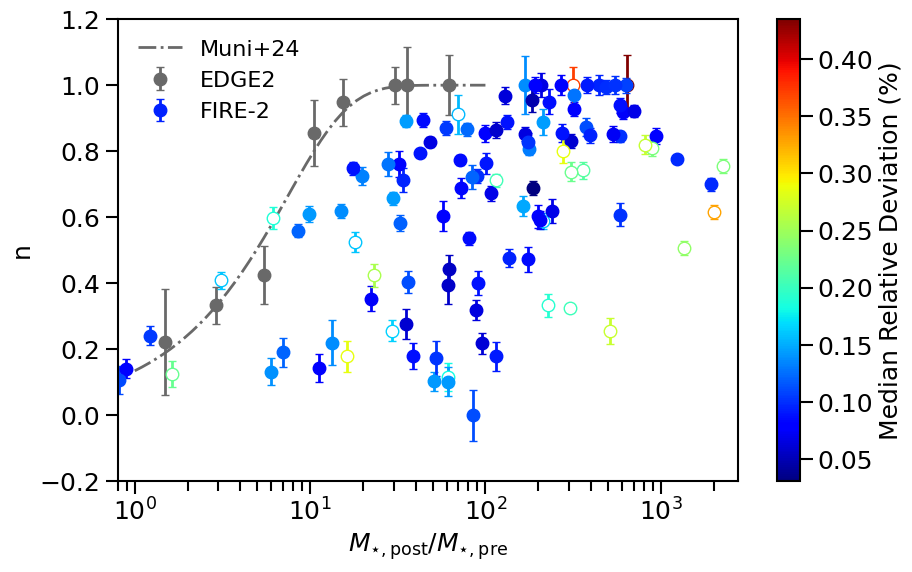}
    \caption{n parameter of the coreNFW profile versus the ratio of stellar mass formed after and before reionisation. Gray circles with errorbars show the results for EDGE2 galaxies and coloured plots indicate the fitted values for galaxies from the FIRE-2 suite. Colours indicate the fit quality via the median deviation of the model relative to the data. Galaxies with median relative deviation over 0.15\% are displayed as empty circles, indicating unreliable fitting around the core radius. The gray dot-dashed line represents the fit to the EDGE2 simulations reported by \cite{Muni_2024}.}
    \label{fig:n_vs_Mpostpre}
\end{figure}

\cite{Muni_2024} also report a tight correlation between the $M_{\star, \rm post}$ / $M_{\star, \rm pre}$ and the inner slope of the dark matter density profile. They illustrate the dependency using the n variable of the coreNFW profile introduced by \cite{coreNFW}. In Fig.~\ref{fig:n_vs_Mpostpre} we compare their findings with the results of fitting the same profile to our dataset of FIRE-2 galaxies. Following their methodology, we fixed the core radius parameter (r$_{\rm c}$) of the coreNFW profile to 1.8 times the three dimensional V-band half-light radius of the galaxy to break the degeneracy with the n parameter when fitting the dark matter density profile. Additionally, we fix the halo mass to the value derived from the simulation and constrain n to lie between 0 and 1. We note that the fixed core radius does not provide satisfactory fits for some galaxies in our sample, which span a broader range of stellar and halo properties than the EDGE simulations. Nevertheless, we confirm via visual inspection that the inner slope of the fitted profile generally matches the data. To quantify the fit quality, we compute the median relative deviation of the model from the data, finding that values above $\sim 0.15$ typically indicate unreliable fits near the core radius. Overall, we find that, due to these fitting limitations, the $n$ parameter is less reliable as a tracer of galactic coreness than the directly measured inner slope of the density profile. We notice FIRE-2 galaxies exhibit lower n values (cuspier) than EDGE galaxies for the same $M_{\star, \rm post}$ / $M_{\star, \rm pre}$, indicating they require more extended SFHs for developing equally shallow density profiles. On top of that, the relation between n and the post- and pre-reionisation stellar mass ratio is much more disperse for FIRE-2 galaxies, which do not follow a clear trend.

\section{Star Formation Histories}
\label{sec-SFHs}

\begin{figure}[ht]
    \centering
    \includegraphics[width=0.5\textwidth]{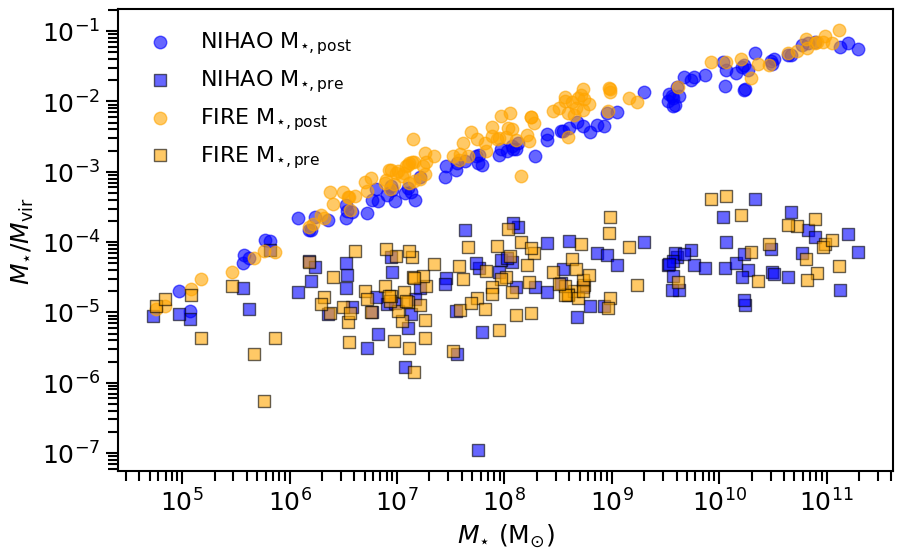}
    \caption{Comparison of pre- and post-reionisation star formation efficiencies. The y-axis shows the stellar mass fractions M$_{\star, pre}$/M$_{\rm vir}$ (squares) and M$_{\star, post}$/M$_{\rm vir}$ (circles) plotted against M$_{\star}$. Data points from the NIHAO and FIRE simulations are highlighted in blue and orange, respectively.}
    \label{fig:mpostpre_mhalo_vs_mstar}
\end{figure}

\mybold{Fig.~\ref{fig:mpostpre_mhalo_vs_mstar} illustrates the stellar-to-halo mass ratios for stellar populations formed both before ($M_{\star, \text{pre}}/M_{\text{vir}}$) and after ($M_{\star, \text{post}}/M_{\text{vir}}$) reionisation, plotted as a function of total stellar mass ($M_{\star}$). In the low-mass regime, post-reionisation star formation is significantly suppressed by stellar feedback. Conversely, pre-reionisation assembly remains relatively unaffected as feedback has not yet had sufficient time to regulate the interstellar medium. This differential impact results in a widening gap between the $M_{\text{post}}/M_{\text{vir}}$ and $M_{\text{pre}}/M_{\text{vir}}$ ratios as galaxy mass increases, fundamentally driving the trend observed in Fig.~\ref{fig:mpostpre_stellarmass_2panels}}

\begin{figure*}[ht]
    \sidecaption
    \centering
    \includegraphics[width=12cm]{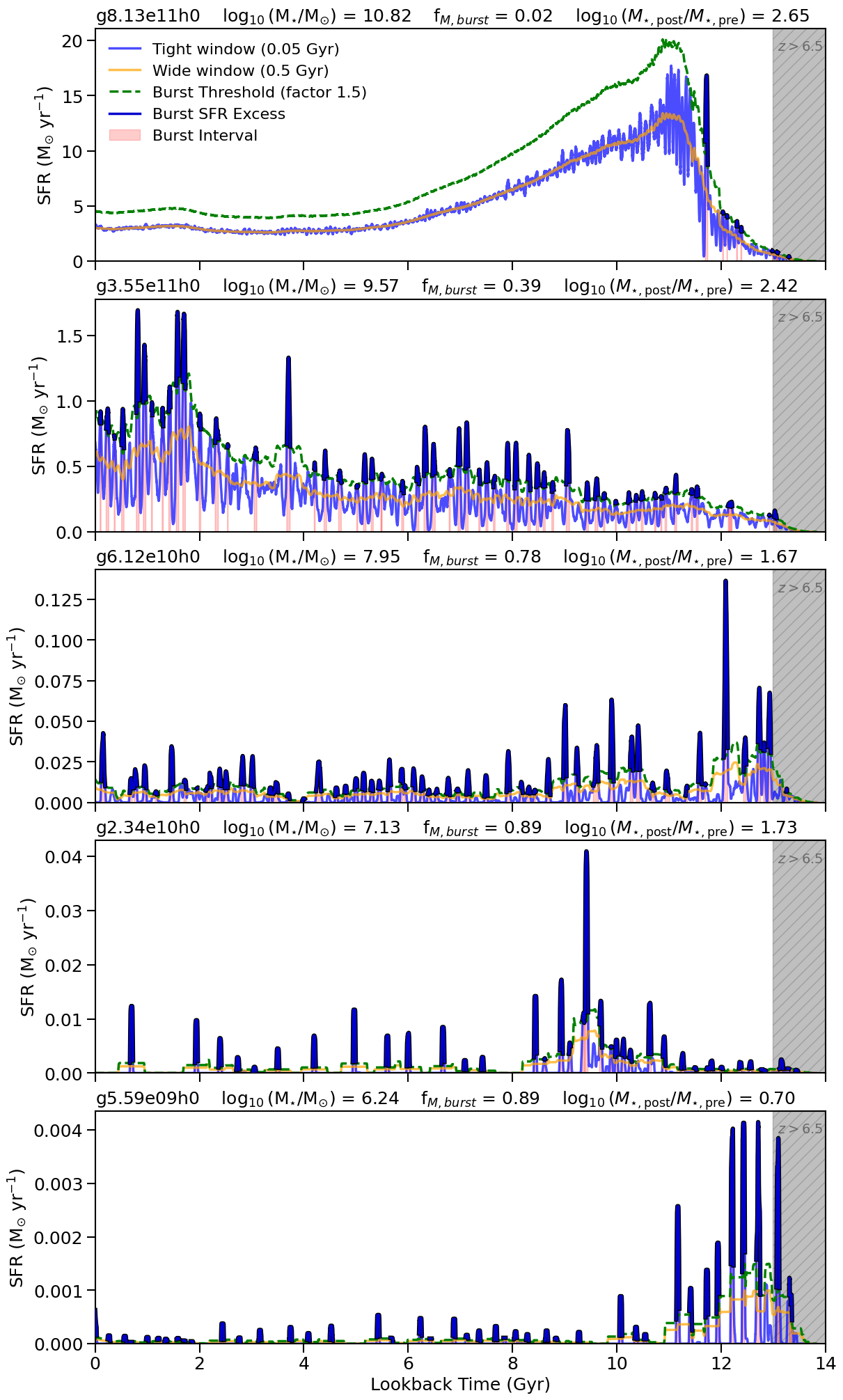}
    \caption{SFR histories for five representative galaxies from the NIHAO simulation suite, sorted in descending order by their stellar mass. Each panel displays the SFR as a function of lookback time for an individual system. Header labels indicate the galaxy identifier, stellar mass, bursty mass fraction, and the mass ratio $M_{\star, \rm post}$ / $M_{\star, \rm pre}$ (refer to Sec. \ref{sec:results-burst} and \ref{sec:results-mpostpre}). The gray shaded region denotes z>6.5, representing the reionisation epoch limit. SFRs are calculated using two sliding temporal windows: 0.5 Gyr (orange) and 0.05 Gyr (blue). A green dashed line indicates a burst threshold, defined as 1.5 times the 0.5 Gyr averaged SFR. High-cadence SFR segments exceeding this threshold are highlighted in dark blue, while light red shaded regions indicate the identified SFR burst intervals.}
    \label{fig:sfr_3_panels}
\end{figure*}

\mybold{In Fig. \ref{fig:sfr_3_panels} we present the SFHs of five galaxies from our NIHAO sample in order to illustrate the calculation process of the bursty mass fraction $f_{\rm M, burst}$ (see Eq. \ref{eq:burst}) and the post-to-pre reionisation stellar mass ratio $M_{\star, \rm post}$ / $M_{\star, \rm pre}$.}

\label{LastPage}
\end{document}